# Goldbach Triples and Key Distribution

Deepthi Cherlopalle and Subhash Kak

**Abstract.** This paper investigates the use of the number of Goldbach triples, or the number of three prime partitions of an odd number, for use in the generation and distribution of cryptographic keys. In addition to presenting randomness properties of these triples, which turn out to be similar to that of prime partitions of even numbers, we explore the question of restricted partition sets. We propose a protocol for key distribution that is based on these numbers. Two of the three partitions of the randomly chosen number serve as cover to send the third number to the two parties that wish to communicate with each other. This third number can serve as session key and the original number of which it is a partition can be used for audit purposes.

## I. INTRODUCTION

Consider the problem of key distribution among n users. For secure communication among these users a total of $n^2$ keys are required and this number increases rapidly as the number of users increases. In the problem of distributing keys in ad hoc networks, the number of users can become uncontrollably large. For two parties to communicate securely over such networks, they must be able to authenticate one another and agree on a secret encryption key. Key establishment protocols are used at the start of a communication session in order to verify the parties' identities and establish a common session key. In key transport protocols, the session key is created by one entity and is securely transmitted to the other. In key agreement protocols, information from both entities is used to derive the shared key.

For some applications efficient scalable one-pass two-party key establishment protocols are needed. In those schemes, only one of the parties transmits information in order to create the session key (but does not transmit the key itself). This means that one-pass approaches lie somewhere between the key transport and key agreement categories. Almost all one-pass approaches belong to the category of authenticated key establishment protocols, because they provide implicit key authentication, meaning that the parties using the protocol are assured that no one else can possibly learn the value of their session key.

Here, we consider the use of Goldbach triples [1] in establishing session keys in a secure protocol. Goldbach numbers as partitions of even numbers were recently proposed for key distribution [2],[3] and the present paper should be seen as a companion paper. The study of these numbers with interesting number theoretic properties is part of an ongoing research program that includes the use of prime reciprocals [4]-[8], Pythagorean triples [9],[10], and permutation transformations [11]-[13]. Number theoretic aspects of partitions are well known in mathematics [14],[15]. The larger problem of cryptography is generally considered from the perspective of computational complexity [16].

The protocol proposed here is a key agreement protocol which is a modification of the one proposed in [2]. The fact that we have three partitions rather than two (if the starting random number is even) provides flexibility that leads to an elegant procedure. We assume that the certification authority (CA) generates the triples of a randomly chosen number and uses the stored values of the hashes of the secret keys of the party as a cover to transport one of the random partitions of the initial number to the two parties. This is done by piggybacking this random number in an XOR sum with the other two triples to the two parties in a manner that the eavesdropper cannot obtain this number even if she had access to the data on both the links between CA and the two parties. The system does not use the secret numbers or keys associated with each user. The knowledge of the hash of the key suffices to authenticate a user. It is assumed that CA has enough controls in place to ensure that the hashes of keys are not compromised. To prevent relay attacks, nonce may also be used.

## II. GOLDBACH TRIPLES

According to the Goldbach weak conjecture, any odd number can be represented (in one or more than one way) as a sum of three primes [1]. The partitioning as three primes is thus a Goldbach triple and it is this number of partitions that will be used to generate random numbers. Some examples:

    9 = (3+3+3)  and (5+2+2)
   11 = (7+2+2) and (5+3+3)
   13 = (7+3+3) and (5+3+3)
   15 = (11+2+2), (7+5+3), (5+5+5)
   17 = (13+2+2), (11+3+3), (7+7+3) and (7+5+5)
   19 = (13+3+3), (11+5+3), (7+7+5)
   21 = (7+7+7), (11+5+5),(11+7+3),(13+5+3) and (17+2+2)

23 = (19+2+2), (17+3+3), (13+5+5), (13+7+3) and (11+7+5)
……

The number of prime partitions for small odd numbers less than 38 is given in Table 1. But as seen in Table 2, the number of partitions increases faster than the number as it becomes large. The number of partitions for 1927 is 3,016 while the number of partitions for 1929 is 2,093.

Table 1.

| Odd Number | Prime Partitions |
|---|---|
| 9 | 2 |
| 11 | 2 |
| 13 | 2 |
| 15 | 3 |
| 17 | 4 |
| 19 | 3 |
| 21 | 5 |
| 23 | 5 |
| 25 | 5 |
| 27 | 7 |
| 29 | 7 |
| 31 | 6 |
| 33 | 9 |
| 35 | 8 |
| 37 | 9 |

## III. STRING GENERATION

We first consider the number of prime partitions of the first 2000 odd numbers. These numbers will be eventually used o generate a set of random numbers. All these numbers are represented as $n$ on the X-axis in Fig. 1 and the Goldbach numbers for these numbers are noted in the corresponding $g(n)$ column along the Y-axis.

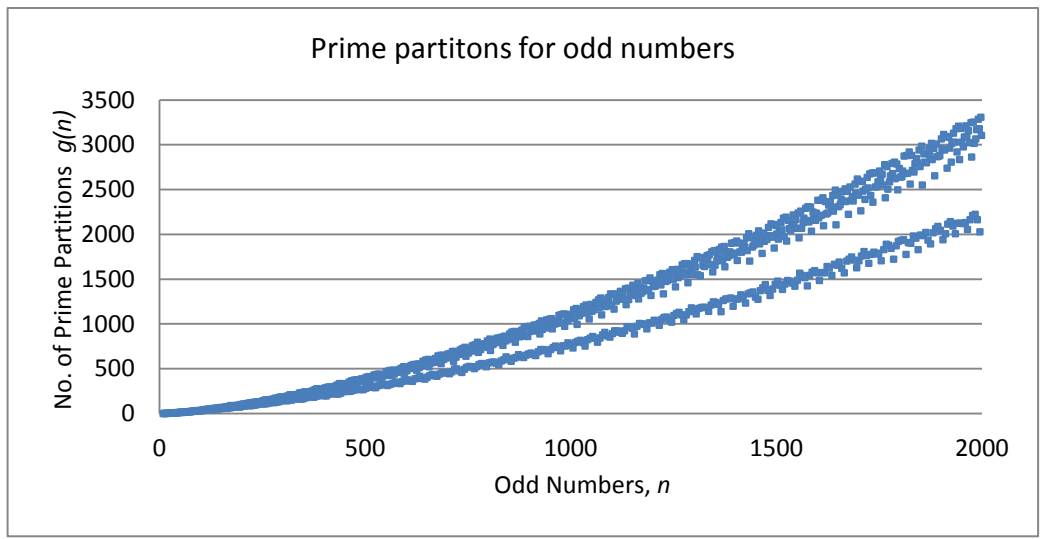

Figure 1. The number of Goldbach partitions

From Table 2, we note that the number of Goldbach triples increases faster than the number. Looking at Figure 1, it is most remarkable that there are two separate ranges of the number of partitions that become clear around $n$ equal to 800 or so. The numbers in Table 2 are thus bunched around 2100 and 3100, respectively.

The following properties are consistent with the data in Table 2:

$g(6k+3) \leq g(6k+7)$ for $k > 10$

g(6k+5) ≥ g(6k+3) for k >10

As for partitions in two primes of even numbers, the function g(n) has local oscillations with a period of 6. Thus there are local minima for n = 171, 177, 183, 189, and so on, and local maxima for n = 173, 179, 185, 191, and so on. As we can see in Table 2, the swing between values becomes increasingly larger as the numbers increase and the values in the Table vary by more than fifty percent.

We now investigate the autocorrelation function of the g(k) sequence. The autocorrelation function captures the correlation of data with itself. For a data sequence X(s), is represented by

$$C(s,t) = corr(X(s), X(t))$$

where, corr is the correlation and X(s) and X(t) are two random variables at points s and t. For a noise sequence, the autocorrelation function will be two-valued, with value of 1 for s=t and a zero value for s≠t. Assuming that we are dealing with ergodic processes, we can, for a discrete sequence, $a_j$, use the following mathematical form of the autocorrelation function:

$$C(k) = \frac{1}{n} \sum_{j=1}^{n} a_j a_{j+k}$$

where n is period and k = 0 to n-1. For computation purposes we map 0 as -1. Figure 2 is the autocorrelation graph for prime partitions (reduced to binary values) of odd numbers up to 1067. For this computation all odd counts were taken to be 1 and all even counts were taken to be -1.

Table 2.

| Odd Number | Prime Partitions |
|---|---|
| 1925 | 2807 |
| 1927 | 3016 |
| 1929 | 2093 |
| 1931 | 3131 |
| 1933 | 3030 |
| 1935 | 2008 |
| 1937 | 3180 |
| 1939 | 2920 |
| 1941 | 2124 |
| 1943 | 3209 |
| 1945 | 2836 |
| 1947 | 2127 |
| 1949 | 3183 |
| 1951 | 3033 |
| 1953 | 2121 |
| 1955 | 2979 |
| 1957 | 3090 |
| 1959 | 2132 |
| 1961 | 3211 |
| 1963 | 3096 |
| 1965 | 2055 |
| 1967 | 3159 |
| 1969 | 3024 |
| 1971 | 2166 |
| 1973 | 3249 |
| 1975 | 2866 |
| 1977 | 2210 |
| 1979 | 3255 |
| 1981 | 3017 |
| 1983 | 2225 |
| 1985 | 3068 |
| 1987 | 3171 |
| 1989 | 2164 |
| 1991 | 3286 |
| 1993 | 3182 |
| 1995 | 2029 |
| 1997 | 3310 |
| 1999 | 3105 |

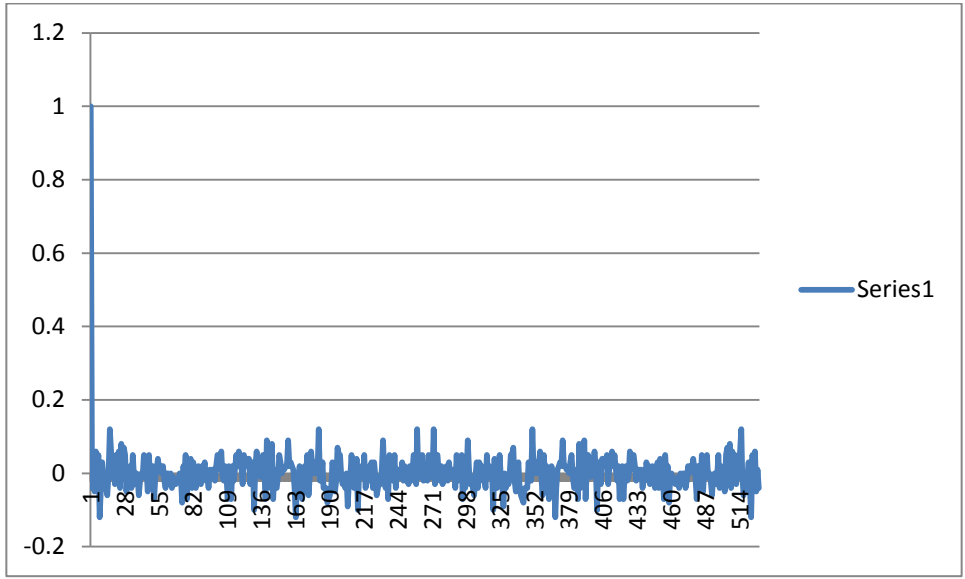
Figure 2. Autocorrelation number of count of Goldbach triples

This autocorrelation function is consistent with that of a pseudorandom sequence.

## IV. TRIANGLE NUMBER SUBSET OF GOLDBACH TRIPLES

In this section we present a further property of number of partitions associated with Goldbach triples of use where we wish to consider a subset of Goldbach triples. Such subsets can increase the cost of the eavesdropper in certain applications.

One specific subset is to consider only such triples that can be valid lengths associated with a triangle. If such numbers are placed as increasing sequence a, b, c, then the largest number c should not be greater than the sum of the other two numbers. We call such numbers triangle numbers, even though this usage has a more restricted meaning in the literature. This property can also be called the geometrical property of "sum of two sides" (the sum of the lengths of any two sides of a triangle must be greater than the third side).

Table 3. Triples satisfying triangularity

| 9=3+3+3 |
|---|
| 11=3+3+5 |
| 13=3+5+5 |
| 15=3+5+7 |
| 15=5+5+5 |
| 17=3+7+7 |
| 17=5+5+7 |
| 19=5+7+7 |
| 21=7+7+7 |
| 23=5+7+11 |
| 25=3+11+11 |
| 25=7+7+11 |

| 27=3+11+13 |
|---|
| 27=5+11+11 |
| 27=7+7+13 |
| 29=3+13+13 |
| 29=5+11+13 |
| 29=7+11+11 |
| 31=5+13+13 |
| 31=7+11+13 |
| 33=7+13+13 |
| 33=11+11+11 |
| 35=5+13+17 |
| 35=7+11+17 |
| 35=11+11+13 |

| 37=3+17+17 |
|---|
| 37=7+13+17 |
| 37=11+13+13 |
| 39=3+17+19 |
| 39=5+17+17 |
| 39=7+13+19 |
| 39=11+11+17 |
| 39=13+13+13 |
| 41=3+19+19 |
| 41=5+17+19 |
| 41=7+17+17 |
| 41=11+11+19 |
| 41=11+13+17 |

| 43=5+19+19 | 47=5+19+23 | 49=13+17+19 |
| 43=7+17+19 | 47=7+17+23 | |
| 43=11+13+19 | 47=11+13+23 | |
| 43=13+13+17 | 47=11+17+19 | |
| 45=7+19+19 | 47=13+17+17 | |
| 45=11+17+17 | 49=3+23+23 | |
| 45=13+13+19 | 49=7+19+23 | |
| | 49=11+19+19 | |
| | 49=13+13+23 | |

We see that these partitions also satisfy properties similar to the unrestricted partitions of odd numbers.

Table 4. Number of triangular partitions

| Odd Number | Triangle Partitions |
|---|---|
| 971 | 232 |
| 973 | 210 |
| 975 | 158 |
| 977 | 244 |
| 979 | 228 |
| 981 | 161 |
| 983 | 251 |
| 985 | 218 |
| 987 | 170 |
| 989 | 260 |
| 991 | 220 |
| 993 | 167 |
| 995 | 233 |
| 997 | 231 |
| 999 | 176 |

For odd numbers from 971 to 999, the number of triangular partitions also falls into two bunches: one around 160 or so and the other around 240 or so.

Table 4 is a count of the first few numbers of prime partitions which satisfy the triangular property. The table also gives the corresponding binary code.

Table 4. Triangular prime partitions and binary code

| Odd Number | Prime Partitions | Binary sequence |
|---|---|---|
| 9 | 1 | 1 |
| 11 | 1 | 1 |
| 13 | 1 | 1 |
| 15 | 2 | -1 |
| 17 | 2 | -1 |
| 19 | 1 | 1 |
| 21 | 1 | 1 |
| 23 | 1 | 1 |
| 25 | 2 | -1 |

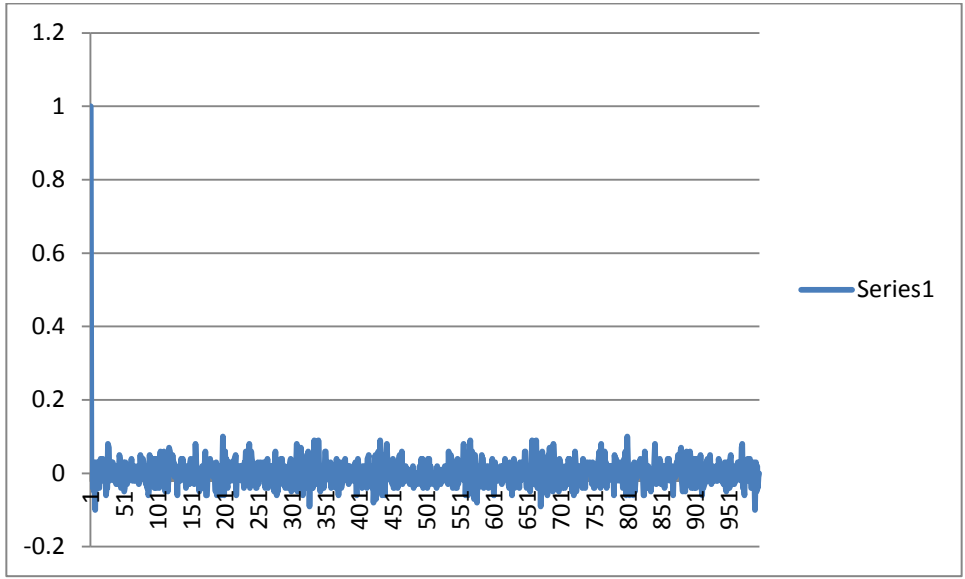
Figure 3. Autocorrelation function for the triangular subset

Figure 3 is the autocorrelation graph generated for input numbers ranging from 1 to 2000 (for the triangular subset). The autocorrelation function in Figure 3 is quite like the function for unrestricted partitions (Figure 2).

## V. GOLDBACH TRIPLES PROTOCOL (GTP)

We now present a protocol for key exchange that uses Goldbach triples. The protocol assumes a Certification Authority (CA) that either issues private keys to Alice and Bob or they choose their own private keys. In the beginning, during the period of registration, Alice and Bob share hashed values of their keys $h(K_a)$ and $h(K_b)$ with the Certification Authority. A cryptographically strong hashing function is used. If such a function outputs number of bits larger than the hashed number of bits for the keys, the extra bit of the hashed value can be ignored.

The private keys of Alice and Bob are $K_a$ and $K_b$, respectively. The protocol is set in motion when one of the parties (say, Alice) approaches CA for a session key for communication with Bob. CA negotiates a connection between Alice and Bob using Goldbach triples. The procedure requires that CA first splits a randomly chosen odd number into a sum of three primes. One of these primes will be the key that will be used by Alice and Bob for their secure communication. The randomly chosen number and its partitions are a part of the audit trail associated with the secure session.

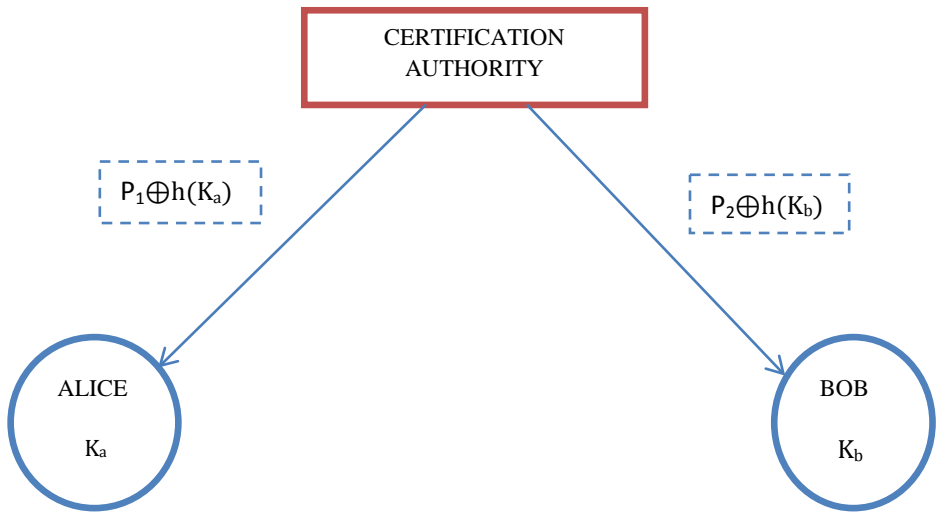
Figure 4. CA sends $P_1$ and $P_2$ under cover of hash functions to Alice and Bob

**Step 1:** CA generates a random odd number (N) as a sum of three primes:

$$N = P_1 \oplus P_2 \oplus P_3$$

CA adds one of the three primes to the hash of the private key (Bit XOR) of Alice and sends it to Alice. Similarly another prime is added to the hash of the private key of Bob and sent to Bob.

In the second part of this Step, Alice and Bob extract $P_1$ and $P_2$ by using their own locally generated hashes of their secret keys, $K_a$ and $K_b$, respectively. This step could also be postponed to the very last as was done in the example that follows.

**Step 2:** CA now sends $P_1 \oplus P_3$ to Alice and $P_2 \oplus P_3$ to Bob. If the eavesdropper has access to both these communications, she will be left with $P_1 \oplus P_2$, which will not reveal any information.

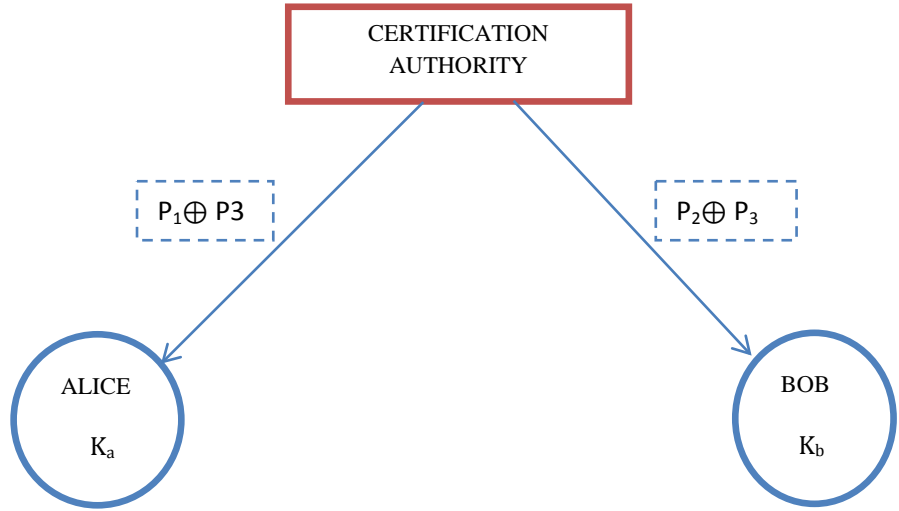

Figure 5. CA sends $P_1 \oplus P_3$ to Alice and $P_2 \oplus P_3$ to Bob

**Step 3:** Both Alice and Bob are now able to extract P3, the random number that will serve as the seed to generate the session key.

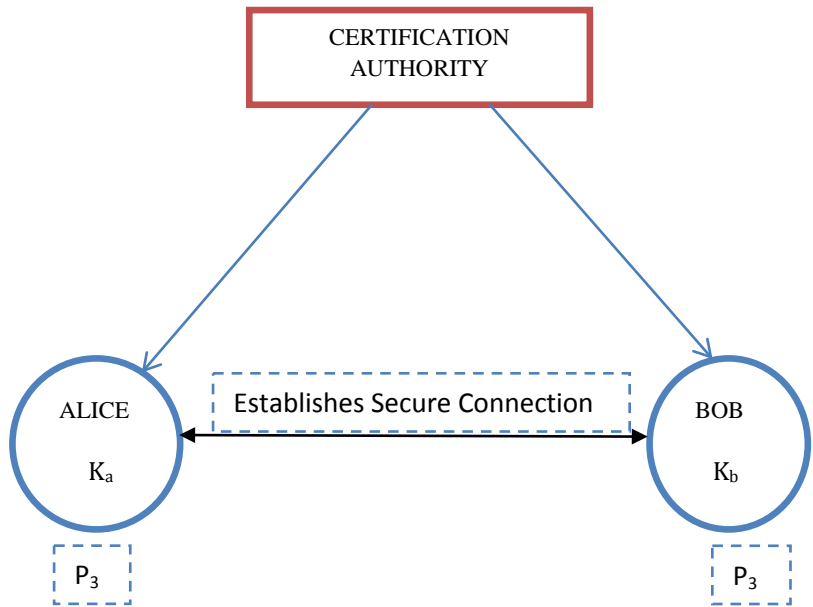

Figure 6. The common key $P_3$ is now used to generate secure link between Alice and Bob

**Example:** Consider the random number N = 181 that has 80 prime partitions.

Let us suppose that CA uses the following partition out of the list: 181=31+67+83. The binary representation of these partitions is as below:

$P_1 = 31 = 0111111$
$P_2 = 67 = 1000011$
$P_3 = 83 = 1010011$

Let $h(K_a) = 47$ and $h(K_b) = 99$

$h(K_a) = 47 = 0101111$
$h(K_b) = 99 = 1100011$

Step 1:

| Alice: $P_1 \oplus h(K_a)$ | Bob: $P_2 \oplus h(K_b)$ |
|---|---|
| 0111111 | 1000011 |
| 0101111 | 1100011 |
| 0010000 → Result1 | 0100000 → Result2 |

Step 2:

| Alice: $P_1 \oplus P3$ | Bob: $P_2 \oplus P3$ |
|---|---|
| 0111111 | 1000011 |
| 1010011 | 1010011 |
| 1101100 → Result3 | 0010000 → Result4 |

| Alice: $P_1 \oplus h(K_a) \oplus P_1 \oplus P3$ | Bob: $P_2 \oplus K_B \oplus P_2 \oplus P3$ |
|---|---|
| 0010000 | 0100000 |
| 1101100 | 0010000 |
| 1111100 → Result5 | 0110000 → Result6 |

Step-3:

| Alice: $h(K_a) \oplus P_1 \oplus h(K_a) \oplus P_1 \oplus P3$ | Bob: $h(K_B) \oplus P_2 \oplus h(K_B) \oplus P_2 \oplus P3$ |
|---|---|
| 0101111 | 1100011 |
| 1111100 | 0110000 |
| 1010011 → Final Key | 1010011 → Final Key |

Finally, Alice and Bob will have obtained the same key ($P_3$) and they establish a secure connection using it. The reason why the session key $P_3$ is sent in the second step, rather than by adding it to the two hashes of the secret keys of Alice and Bob in the first step itself, is to protect the hashes of the keys in case the session key gets compromised. To this extent, this protocol is an improvement on the version that was presented in [3].

## VI. CONCLUSIONS

This paper investigated the use of Goldbach triples for use in the generation and distribution of cryptographic keys. We presented randomness properties of these triples and explored the question of restricted partition sets. We proposed a protocol for key distribution that is based on Goldbach triples that appears to provide greater security than another protocol based on two prime partitions [3]. Two of the three partitions of the randomly chosen number serve as cover to send the third number to the two parties that wish to communicate with each other. This third number can serve as session key and the original number of which it is a partition can be used for audit purposes.

**Acknowledgement.** This research was supported in part by research grant #1117068 from the National Science Foundation.